# Securing Cluster-heads in Wireless Sensor Networks by a Hybrid Intrusion Detection System Based on Data Mining

Mohammadreza Darabi
*Tolue inc., Tehran, Iran*
*Email: m.darabi@tolue.com*

*Abstract*- Cluster-based Wireless Sensor Network (CWSN) is a kind of WSNs that because of avoiding long distance communications, preserve the energy of nodes and so is attractive for related applications. The criticality of most applications of WSNs and also their unattended nature, makes sensor nodes often susceptible to many types of attacks. Based on this fact, it is clear that cluster heads (CHs) are the most attacked targets by attackers, and also according to their critical operations in CWSNs, their compromise and control by an attacker will disrupt the entire cluster and sometimes the entire network, so their security needs more attentiveness and must be ensured. In this paper, we introduce a hybrid Intrusion Detection System (HIDS) for securing CHs, to take advantages of both anomaly-based and misuse-based detection methods, that is high detection and low false alarm rate. Also by using a novel preprocessing model, significantly reduces the computational and memory complexities of the proposed IDS, and finally allows the use of the clustering algorithms for it. The simulation results show that the proposed IDS in comparison to existing works, which often have high computational and memory complexities, can be as an effective and lightweight IDS for securing CHs.

*Index Terms*- Wireless Sensor Networks (WSNs), Intrusion Detection Systems (IDSs), Cluster-heads (CHs), data preprocessing model, clustering algorithms.

I. INTRODUCTION

Wireless Sensor Networks because of their inherent advantages such as lower cost and easier deployment on the environment, to play a role in a wide range of applications such as military surveillance [1], fire control in forest, health care [2], safety monitoring on Structures and buildings, and smart homes [3] are highly desirable and cost-effective. However, resource constrains, such as limited processing power, memory and energy are main challenge in WSN design and application [4].

Given that WSNs are often used in remote and unprotected locations or where adverse operating conditions or even hostile operating conditions, they are highly susceptible to intrusions and security attacks [5]. Most of attacks try to cause a sharp decline in network performance using this weakness. Therefore, security in WSNs has become an important issue, especially if these networks are involved in critical processes. Secure WSNs have critical importance in the military (tactical) applications, so that a security gap in the network can weaken its own forces on the battlefield [6].

Also it should be noted that in any secure or less secure network, it cannot be completely prevented from intrusions. When attacking to a network and intrusion to it, some nodes are captured by the attacker and thus malicious node can identify and reveal their confidential information such as security keys. This will lead to the failure of the intrusion prevention operation and Jeopardize network security. In such a situation, the existence of an Intrusion Detection System (IDS) in the network by timely detecting of intrusions can prevent the disclosure of security information and the waste of resources.

In the other words, by deploying IDS that is a set of tools, methods, and resources to help identify, assess, and report intrusions in these networks, we can keep the network efficiency at the optimum level by detecting timely attacks, and the protecting the network from security threats.

One of the challenges of using IDSs in CWSNs is securing CHs. Since CHs are of great importance in WSNs and perform operations of cluster management, data aggregation, and data transfer to the base station, they are much more likely to be attacked than normal nodes, Such that the intrusion and control of a CH by an attacker will disrupt the entire cluster operation and in some cases the entire sensor network. So, in a sensor network, maintaining the security of CH nodes and in some way guaranteeing it is very important. On the other hand, the use of IDSs for common nodes, such as proposed IDS in [7], is not suitable for CH nodes, due to their high sensitivity and the need for security guaranty.

The two main methods of intrusion detection are anomaly-based detection and misuse-based detection, which none of them alone can provide security for CH nodes. The anomaly-based intrusion detection method has a high detection rate, but its disadvantage is the high rate of false alarms. On the other hand, the misuse-based detection method has high accuracy in detecting attacks, and a low rate of false alarms, but unfortunately its detection rate is relatively low. Therefore, the best option for securing CHs is to use a hybrid intrusion detection method, which, due to its high computational complexity, leads to an increase in energy consumption, which makes a problem for the sensor network. Of course, in most cases, the CH nodes according to their respective operations have higher capabilities than the common nodes, which allows us to use the more effective IDSs with respect to high-security sensitivity of them. But as much as possible, the computational overhead and, consequently, energy consumption should be reduced in order to increase the lifetime of the network.

In this paper, we present a hybrid IDS based on data mining algorithms for securing CHs, which by using a novel data pre-processing model reduces the computational complexity and consumption memory in the IDS, and it allows us to use the data mining classification algorithms for detect intrusions and securing CHs in WSNs. Therefore, in the proposed system, in addition to the benefits of both anomaly-based detection and misuse-based detection methods, which led to a high detection rate and low false alarms rate, with the help of the proposed novel data pre-processing model, energy consumption will be at least, which is very important in WSNs.

In order to evaluate and present the results of the proposed method, and also because of the absence of a real sample of the dataset for intrusion detection in WSNs, the KDDCup'99 dataset is used as the sample to evaluate the performance of IDSs in these networks. The simulation results show that the proposed IDS in comparison to existing works, which often have high computational and memory complexities, can be as an effective and lightweight IDS for securing CHs.

This paper is organized as follows: In Section II, we introduce the IDSs and then the Dataset for IDSs are described. In Section III, a review on the most important IDSs devised for WSNs is presented along with the introduction of their advantages and shortcomings. Section IV describes the proposed IDS. In Section V, we will simulate the proposed IDS and present the related results. Finally in Section VI, the paper ends with a conclusion and future works.

## II. PRELIMINARIES

In this section, IDSs are described along with their types and requirements, and then Datasets for IDSs are introduced.

### *A. Intrusion Detection Systems*

In general, any type of unauthorized or unwanted activity in a network is called intrusion. An IDS is a set of tools, methods, and resources to help identify, assess, and report intrusions. IDS is not a single, separate unit, but rather part of an overall protection system that is installed alongside a network node. Intrusion is defined as any set of activities that attempt to endanger the integrity, confidentiality or availability of a resource, and Intrusion Prevention System (IPS) includes methods such as encryption, authentication, key management [8], [9], access control, secure routing, etc. is considered as the first line of defense against intrusions [10].

However, it should be noted that in any secure or less secure network, IPS cannot be completely prevented from intrusions. Therefore, after IPSs, IDSs are considered as the second line of defense against attacks and intrusions. The expected operating conditions in IDS will be as follows [11], [12]:

- Not add new flaws and weaknesses to the network.
- Less use of network resources, and not reducing performance by imposing overhead.
- Low False alarm rate, which is the percentage of normal activity that is detected as anomaly.

- High detection rate, which is the percentage of anomalies that have been properly detected.
- Run continuously and act impalpable for the system and users (Transparency principle).
- Should be in accordance with standards to allow for future cooperation and development.

Each IDS has three main components [12], [13]:

- *Monitoring Section:* This section is used to monitor local events and neighbors and often by traffic analysis and local events, controls the resources efficiency.
- *Analysis and Detection:* This module is the main part of the IDS, which is dependent on the modeling algorithm. In this section, the behavior and activities of the network are analyzed and decided to declare them as an intrusion.
- *Warning section:* This section is responsible for reaction against intrusion, which generates an alarm about the detection of an intrusion.

IDSs are categorized into three groups based on their operation, which are described below [10], [14]:

**Anomaly-based Detection:** This method is based on a statistical behavior model related to the normal operations of network nodes that are profiled and if there is a certain deviation from it, as an anomaly is detected. In the other words, this method first describes the actual features of a 'normal behavior', and then detects any activities that deviate from these behaviors as intrusions. The main advantage of this method is its high detection rate, but on the other hand, its disadvantage is also that it generated a high false alarms rate.

**Misuse-based Detection:** In this method, the patterns of previously known attacks are produced and used as a reference for identifying future attacks. The advantage of this technique is that it can accurately and efficiently detect known attacks. The disadvantages are that this technique needs knowledge to build attack patterns and they are not able to detect novel attacks. So this method has a low false alarms rate, but its detection rate is also relatively low.

**Specification-based detection:** This method defines a set of specifications and constraints that describe the correct operation of a program or protocol. Then the program execution is monitored according to the defined specifications and constraints. In fact, this method combines the aims of misuse and anomaly detection methods, that is able to detect previously known attacks at low false alarms rate. The disadvantage of this method is the manual setting of all specifications, which is a time-consuming process for users.

*B. Dataset for Intrusion Detection Systems*

Because of the absence of a real sample of the dataset for intrusion detection in WSNs, the KDDCup'99 dataset is used as the sample to evaluate the performance of IDSs in these networks. The KDDCup'99 dataset was designed by Columbia University through the simulation of intrusions and attacks in a military network environment at the DARPA organization in 1998 [15]. It was performed in the MIT Lincoln Research Labs, and then announced on the UCI KDD Cup 1999 Archive. Each sample of this dataset represents a connection between two network hosts according to network protocols and is described by 41 features that consist of 34 types of numerical features and 7 types of symbolic features.

TABLE I. Four classes in KDDCup'99 dataset with their description and type of attacks.

| # | Class Type | Description | Attack Type |
|---|---|---|---|
| 1 | DOS | In DoS, an attacker tries to prevent legitimate users accessing or consume a service.<br>Select Forward, which uses illegitimate data forwarding to make an attack, is known as a DoS attack. | Smurf, Back, Neptune, Teardrop, Land, Pod. |
| 2 | Probe | In Probe attack, an attacker tries to gain information about the victim machine. The intention is to check vulnerability on the victim machine. e.g. Port scanning. The attacks of Spoofed, Altered, or Replayed Routing Information, Sinkhole, Sybil, Wormholes, and Acknowledgment Spoofing need to make a probe step before they begin to attack, so they would be classified as Probe attacks. | Portsweep, Satan, Ipsweep, Nmap. |
| 3 | R2L | The attacker tries to gain access to the victim system by compromising the security via password guessing or breaking.<br>Spoofed, Altered, or Replayed Routing Information, Sinkhole, Sybil, Wormholes, Hello Floods, and Acknowledgment Spoofing use the weakness in the system to make an attack, so they would be classified as R2L. | Buffer_Overflow, Guess_passwd, Warezclient, Spy, Warezmaster, Phf, Multihop, Imap. |
| 4 | U2R | In U2R, an attacker has local access privilege to the victim machine and tries to access super users (administrators) privileges via "Buffer overflow" attack.<br>Sinkhole, Wormholes, and Hello Floods are caused by inner attacks, and are therefore classified as U2R. | Loadmodule, Perl, Ftp_write, Rootkit. |

All Features can be classified into four different classes as discussed below [16]:

- Basic Features are the attributes of individual TCP connections.
- Content features are the attributes within a connection suggested by the domain knowledge.
- Traffic features are the attributes computed using a two-second time window.
- Host features are the attributes designed to assess attacks which last for more than two seconds.

Each sample is labeled as either a normal behavior or one specific attack. The dataset contains 23 class labels that one is normal and the remaining 22 are different attacks that are categorized into four classes: DoS, Probe, R2L, and U2R. In Table I, these four classes are presented with their description and type of attacks [15], [17].

In this paper, we used kddcup.data_10_percent.gz as our sampling step in creating the training and testing datasets. This dataset contains 10% data in KDDCup'99 dataset, where the total number of sample records is 494,021. The complete statistics for this dataset are presented in Table III.

III. RELATED WORK

So far, many IDSs have been introduced for WSNs, but there is still competition for increasing the detection rate, reducing the false alarms rate and minimizing energy consumption. Considering high sensitivity and the need for security guaranty in CHs, and Also disadvantages of anomaly-based detection and misuse-based detection, none of them alone is capable of securing CH nodes. Therefore, the best option for securing CHs is to use a hybrid IDS, which has been proposed in many references [18]-[28]. In the following, we will introduce the most important related works.

In [18], a hybrid Intrusion Detection System is proposed for Cluster-based WSNs that detect malicious nodes by integrating misuse detection rules and functional reputation. The main idea of the proposed method is that instead of detecting attacks only at nodes level, they propose a collaborative and centralized design using the mutual trust assessment between all network components, in which each sensor node computes functional reputation values for its neighbors by observing their activities (transmissions and data aggregation). In order to achieve this, they have defined five functional reputation metrics and benefit from the high detection rate of the misuse detection method by applying the relevant rules. The main problem with their methodology is that only have expressed their energy consumption results and have not presented any discussion of the detectable types of attacks and their detection rates.

In [19], an Integrated Intrusion Detection System (IIDS) is proposed in a heterogeneous Cluster-based WSN. According to the different capabilities and probabilities of attacks on them, three separate IDSs are designed for the sink, CH and Sensor Node (SN). For CHs, a Hybrid IDS is proposed, which combines anomaly and misuse detection. They reduced the number of features using the SVM method to 24 features, and finally use a three-layer Back-Propagation Network (BPN) for classification. Their IDS, due to the low false alarms rate and also low computational complexity, can be used in WSNs, but the main problem is the relatively low detection rate, given the importance of CHs.

In [20], a Global Hybrid IDS (GHIDS) has been proposed that to achieve the goal of high detection rates and low false alarms, used combination of a technique based on support vector machine (SVM) for detecting anomalies, with a set of signature-based detection rules to identify attacks in cluster-based WSNs. The results of the simulations show that the proposed method is in a desirable condition In terms of the detection rate and the false alarm rate. But the underlying problem is the high energy consumption due to the use of an anomaly detection technique based on SVM, which is somewhat inappropriate for the sensor network.

In [21], a similar method with [20] has been proposed, that to reduce the computational complexity and energy consumption, existing features reduced to 4 features. Therefore, a significant improvement has been created in energy consumption, but its detection rate is proportionally lower.

In [22], [23] and [24], hybrid IDSs have been proposed that initially use a novel algorithm to feature selection in order to reduce the computational complexity, and then use the SVM algorithm for classification. In [22] uses the combination of ant colony optimization and a feature weighting SVM to effective feature selection that finally reduces the number of features to 25. In [23] uses GA to feature selection that finally reduces the number of features to 10. In [24] also uses the intelligent water drops (IWD) algorithm, a nature-inspired optimization algorithm for feature selection that finally reduces the number of features to 9. The main problem of all three methods is the relatively high computational complexity due to the use of the SVM classification algorithm.

In [25], a Modified CuttleFish Algorithm (MCFA) approach is proposed that plays a crucial role in intrusion detection by selecting an appropriate subset of the most relevant features from the huge amount of dataset. Griewank fitness function is used to calculate the fitness of the MCFA. Naïve-Bayes classifier is also employed as a classification algorithm.

In [26], an entropy-based feature selection to select the important features, layered fuzzy control language to generate fuzzy rules, and layered classifier to detect various network attacks is proposed.

In [27], an improved many-objective optimization algorithm (I-NSGA-III) is proposed using a novel niche preservation procedure. It consists of a bias-selection process that selects the individual with the fewest selected features and a fit-selection process that selects the individual with the maximum sum weight of its objectives. Experimental results show that I-NSGA-III can alleviate the imbalance problem with higher classification accuracy for classes having fewer instances.

In [28], a Knowledge-Based intrusion Detection Strategy (KBIDS) is proposed to detect several types of attacks under different network structures, that aims to create a stand-alone detection model from network structure for WSNs. Their proposed mechanism is based on the fact that various types of attacks are very likely to have various forms of density in the feature space. They collected the network traffic and used it as the characteristics of the behavior of random networks in the feature space. Then the density forms can be considered as an indicator for detecting normal and abnormal network behavior. The simulation results of the proposed method in [28] on the sinkhole, hello flooding and DoS attacks indicate the proper detection accuracy and high compatibility with the network structure than the existing works.

IV. PROPOSED INTRUSION DETECTION SYSTEM

One of the challenges of using IDSs in cluster-based WSNs is securing CHs. Since CHs are of great importance in WSNs and perform operations of cluster management, data aggregation, and data transfer to the base station, they are much more likely to be attacked than normal nodes, Such that the intrusion and control of a CH by an attacker will disrupt the entire cluster operation and in some cases the entire sensor network. So, in a sensor network, maintaining the security of CH nodes and in some way guaranteeing it is very important. On the other hand, the use of IDSs for common nodes, such as proposed IDS in [7], is not suitable for CH nodes, due to their high sensitivity and the need for security guaranty. Also, due to the disadvantages of anomaly-based detection and misuse-based detection, none alone is capable of securing CH nodes.

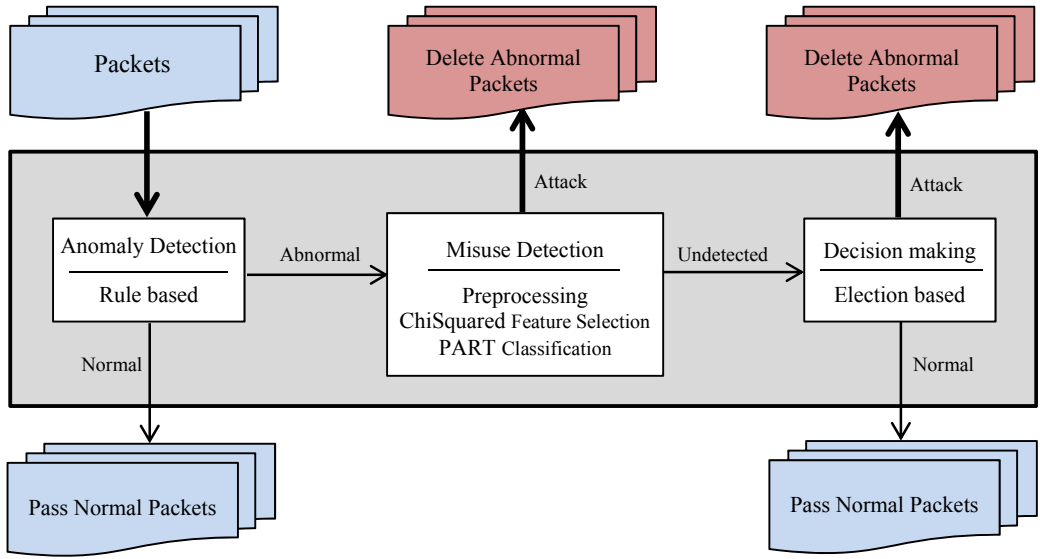

Fig. 1. Operations of proposed intrusion detection system.

On the other hand, considering that energy is as a critical parameter in WSNs, and practically the network lifetime depends on it, a lightweight method should be used for intrusion detection in them. Of course, in most cases, the CH nodes have more capabilities than common nodes due to their respective operations. Therefore, in order to secure CHs, with respect to their high-security sensitivity, we can use the more efficient IDSs.

In this section, we present a hybrid IDS based on data mining algorithms for securing CHs, which by using a data pre-processing model reduces the computational complexity and consumption memory in the IDS, and it allows us to use the data mining classification algorithms for detect intrusions and securing CHs in WSNs. Therefore, in the proposed system, in addition to the benefits of both anomaly-based detection and misuse-based detection methods, which led to a high detection rate and low false alarms rate, with the help of the proposed data pre-processing model, energy consumption will be at least, which is very important in WSNs.

As shown in Fig. 1, the process of the proposed IDS is such that initially received packets from other nodes are examined by anomaly detection model. Anomaly detection model (described in Section IV-A) can quickly filter large numbers of normal packets and then deliver abnormal packets to the misuse detection model (described in Section IV-B) to identify attacks and their types there. Finally, packets that are not detected by a misuse detection model will also be identified at the decision-making step. In the following, we will describe the details of each step of the proposed IDS.

*A. Proposed Anomaly-based detection model*

The anomaly detection model is used as the first line of defense in the proposed IDS. Given that a large number of existing packets, in fact, only a few of them are related to attacks, and most of them are also related to the network normal state, so using an abnormal detection model that acts like a filter, Quickly, the normal packets are passed and the abnormal packets are filtered and delivered to the misuse detection model to more detecting and accurately. An anomaly detection system uses a defined model of network normal behavior, so a packet is detected as an anomaly by the system when the current behavior deviates in comparison with the defined behavior.

TABLE II. Rules in the anomaly detection model for attacks detection in WSNs.

| # | Rule | Description | Detectable attacks |
|---|---|---|---|
| 1 | Interval Rule | Time between reception of two consecutive messages exceeds lower and upper limits | Denial-of-service attacks Hello flood attack |
| 2 | Retransmission rule | A message should be forwarded by the middle nodes to other nodes. | Sinkhole and selective forwarding attacks |
| 3 | Integrity rule | The original message should not be changed along the path in between the source and destination nodes. | Content modification attack |
| 4 | Delay rule | A retransmission of a message should be done after a certain waiting time. | Denial-of-service attacks |
| 5 | Repetition rule | Number of retransmissions of a message by a neighbor node exceeds limit | Denial-of-service attacks |
| 6 | Radio transmission range rule | A single message should be received from neighboring nodes, which can be identified by the RSSI. | Sybil, Wormhole and Hello flood attacks |
| 7 | Jamming rule | Number of collisions of message sent by monitor node exceeds expected limit | Jamming attack |

One of the problems with anomaly detection model is that if the current behavior and normal behavior patterns change in the network, then the system usually detects the normal communication as abnormal communication and creates the problem of erroneous classification. However, it rarely detects abnormal communication as normal communication.

In order to solve the erroneous classification problem in the anomaly detection model, in the second line of defense, we use a misuse detection system to take delivery the detected abnormal packets by the anomaly detection model and, with more accurate analyzes, Their final status will be determined. in other words, the abnormal detection model, with the receipt a large number of packets, the abnormal cases that are relatively few, like a filter separates from a large number of normal ones, and after passing normal packets in high accurately, the abnormal cases for More accurate examination are delivered to misuse detection model.

As mentioned, to create an anomaly detection model to monitor the status of data packets, normal behavior patterns must be created in the network, which in this paper, due to requiring the high performance; a rule-based analysis method is used. According to reference [29], the rules of Table II are considered in the rule-based method to create an anomaly detection model.

*B. Proposed misuse-based detection model*

The misuse Detection Module uses various models of known attack behavior, so we need to create a basic model that matches these behaviors. Because the performance in most IDSs is guaranteed through training data, machine learning methods are consistent with this approach.

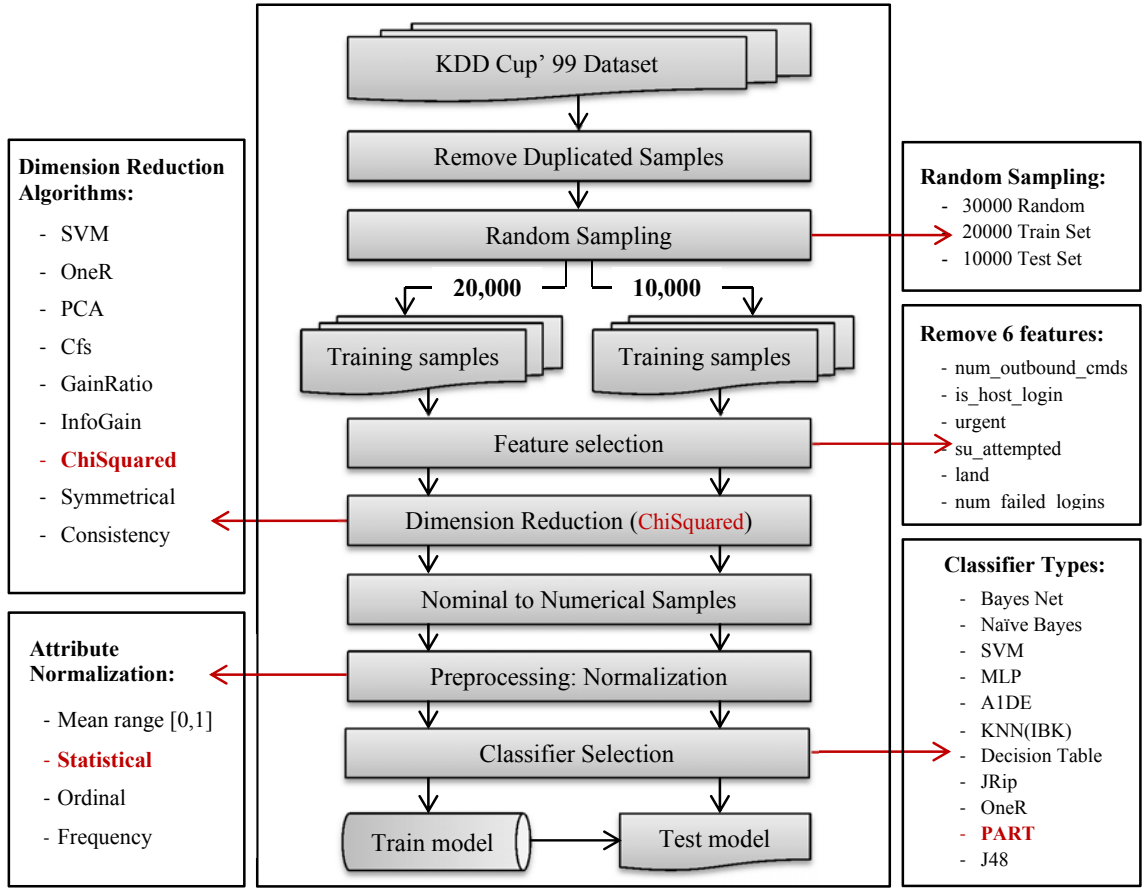

Fig. 2. Steps of the proposed data pre-processing model for the dataset.

In this section, we present a data pre-processing model for increasing the efficiency of the IDS as well as reducing the energy consumption, and finally, to obtain the best detection rate, different machine learning algorithms are examined for classification. Fig. 2 shows the steps of the proposed pre-processing model for the dataset.

One of the effective factors in increasing the computational complexity and memory consumption in using data mining methods is the number of training samples for creating the model. Therefore, given the large number of samples in the dataset, it is practically impossible to use it in WSNs. So, in the first and second steps of the proposed model, we try to use the techniques to optimize the number of available samples to usability in WSNs.

**1. Remove duplicated samples:** As shown in Fig. 2, in view of the data redundancy in the data set, we initially do the remove operation of duplicated samples. As shown in Table III, by removing the duplicated samples, the size of data sets decreases sharply (reduction rate 70.53%), which, in addition to reducing computational complexity and reducing energy consumption, causing to increased detection accuracy and reduced memory consumption.

**2. Random sampling:** after remove duplicated samples, using random sampling of all records in the dataset, we sample 20,000 records as training data and 10,000 records as testing data. Given that the sample set of Probe, U2R, and R2L attacks is very small; hence, the whole their records are sampled, So that two-thirds of these records are taken as training data and one-third as testing data; but other sample sets are selected according to their ratio from kddcup.data_10_percent.gz dataset that detailed in Table III.

TABLE III. Number of samples and its ratio in the KDDCup'99 dataset.

| Category | Total data | | No Duplicated data | | Training data | | Testing data | |
|---|---|---|---|---|---|---|---|---|
| | samples | Ratio(%) | samples | Ratio(%) | samples | Ratio(%) | samples | Ratio(%) |
| Normal | 97278 | 19.69% | 87832 | 60.33% | 11079 | 55.40% | 5549 | 55.49% |
| Dos | 391458 | 79.24% | 54572 | 37.48% | 6798 | 33.99% | 3393 | 33.93% |
| Probe | 4107 | 0.83% | 2130 | 1.46% | 1421 | 7.11% | 709 | 7.09% |
| R2L | 1126 | 0.23% | 999 | 0.69% | 667 | 3.33% | 332 | 3.32% |
| U2R | 52 | 0.01% | 52 | 0.04% | 35 | 0.17% | 17 | 0.17% |
| **TOTAL** | **494021** | **100%** | **145585** | **100%** | **20000** | **100%** | **10000** | **100%** |

TABLE IV. 6 features with the least importance and no distinction from the KDDCup'99 dataset.

| is_host_login | num_outbound_cmds | urgent | su_attempted | land | num_failed_logins |
|---|---|---|---|---|---|

Considering that a large number of features are also one of the most important factors in increasing the computational complexity and energy dissipation, as well as significantly increase the memory consumption, practically using of data mining methods in WSNs according to the computational and memory constraints of their nodes make it impossible. Therefore, in order to overcome this problem and reduce the computational complexity and energy dissipation as well as memory consumption, we must use techniques to reduce the number of features to the appropriate number, which is also done in steps 3 and 4 in the proposed model.

**3. Feature Selection (deletion of ineffective features):** In order to optimize the dataset in the first step, with a superficial observation, it is easy to select several attributes due to the lack of a distinction in the dataset and to remove them from the dataset. As shown in Figure 2, this step is presented as the feature selection, in which 6 features with the least importance and no distinction from the dataset are eliminated. For example, the is_host_login and num_outbound_cmds features in the entire dataset records are zero and therefore do not create any distinction in the datasets. These features are presented in Table IV.

**4. Dimension reduction and selection of effective features:** In order to further reduce the computational complexity and energy dissipation in the WSN nodes, we use a feature selection algorithm to reduce the dimension in the dataset.

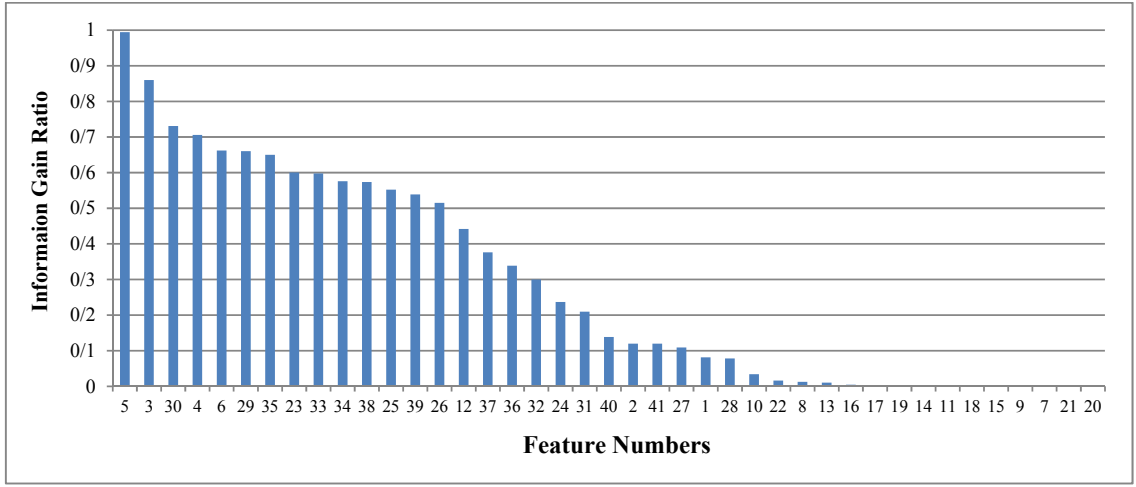

Fig. 3. Rating of 41 features in the KDDCup'99 dataset based on the information gain ratio (IGR).

Among the features in the dataset, they all have no deterministic effect on the output, and even some of them increase the classification error. In Fig. 3, the rating of 41 features in the KDDCup'99 dataset is presented based on the information gain ratio (IGR) that sorted in a descending order. As seen in the figure, most of the features have an IGR under the average of the dataset (IGR average = 0.29). In fact, only 20 features are above the average, which indicates that the original dataset is concentrated in a small group of values.

The features that lead to the convergence of connection categories within a small group of values have very little information to describe the behavior of a node in the network. This indicates that the original dataset contains a series of irrelevant data for the IDS and so needs to be optimized.

Therefore, feature selection is an important step in the optimization of the dataset that can have a desirable effect on the performance of the IDS. In order to select an effective set of features, we examined the most important methods for feature selection. In Table V, we presented the results based on the detection rate of different classification algorithms.

As seen in Table V, the most reduction of features is related to the ChiSquared method with four selected features, which however with a high detection rate of 99.59%, has very desirable conditions for use in WSNs. Also, the InfoGain method with 11 features and the detection rate of 99.72, has the ability to use in WSNs, but due to about 3 times the selected features in comparison with the ChiSquared method, a higher computational overhead, and as a result, higher energy consumption imposes to the system. So, in this paper, we use the ChiSquared feature selection method to dimension reduction of the dataset. The four selected features are presented in Table VI for increasing the efficiency of the proposed IDS.

**5. Data normalization:** In the last step, we will normalize the dataset. In reference [30], the normalization of features has been considered as an essential step in the data preprocessing in order to

13TABLE V. Comparison of various Feature Selection Methods based on Detection Rate in KDDCup'99 dataset.

| Feature Selection Approaches | Selected Features | Detection Rate of Various Classifiers | | | | | | | | |
|---|---|---|---|---|---|---|---|---|---|---|
| | | Random Tree | J48 | Bayes Net | **PART** | JRip | Random Forest | A1DE | Decision Table | Naïve Bayes |
| Full Features | 41 | 99.42 | 99.53 | 96.54 | 99.64 | 99.69 | **99.80** | 99.80 | 99.17 | 86.01 |
| Remove ineffective Features | 35 | 99.38 | 99.48 | 96.44 | 99.57 | 99.62 | **99.80** | 99.78 | 99.08 | 87.18 |
| **Chi Squared** | 4 | 99.40 | 99.32 | 97.97 | **99.59** | 99.45 | 99.44 | 99.49 | 98.34 | 89.57 |
| One R | 7 | 99.43 | 99.42 | 97.41 | 99.51 | **99.57** | 99.54 | 99.53 | 98.41 | 91.11 |
| Consistency Subset + BFS | 8 | 99.45 | 99.6 | 97.62 | 99.58 | 99.59 | 99.55 | **99.61** | 99.06 | 82.62 |
| Info Gain | 11 | 99.45 | 99.52 | 96.54 | 99.58 | 99.60 | **99.72** | 99.71 | 99.06 | 88.69 |
| CFS Subset + BFS | 13 | 99.39 | 99.48 | 97.01 | 99.48 | 99.56 | 99.63 | **99.69** | 98.97 | 91.99 |
| Symmetrical Uncert | 15 | 99.39 | 99.55 | 96.37 | 99.55 | 99.64 | **99.76** | 99.67 | 99.10 | 86.75 |
| Correlation | 15 | 98.79 | 98.91 | 93.62 | 99.01 | 98.96 | **99.24** | 98.36 | 96.52 | 81.61 |
| ReliefF | 15 | 98.54 | 98.72 | 93.91 | 98.78 | 98.66 | **99.01** | 98.62 | 96.75 | 89.97 |
| Gain Ratio | 16 | 99.43 | **99.66** | 97.30 | 99.61 | 99.59 | 99.60 | 99.56 | 98.83 | 88.45 |
| SVM | 22 | 99.56 | 99.56 | 96.42 | 99.64 | 99.70 | **99.78** | 99.75 | 99.05 | 84.94 |

TABLE VI.: Selected features with ChiSquared feature selection algorithm.

| Feature # | Feature Name | Description |
|---|---|---|
| 3 | service | service on the destination, e.g., http, telnet,etc. |
| 5 | src_bytes | Number of data bytes from source to destination |
| 30 | diff_srv_rate | % of connections to different services |
| 35 | dst_host_diff_srv_rate | Dif_srv_rate for destination host |

increase the efficiency of IDSs. In this reference, four different schemes for the normalization of features at the preprocessing stage have been introduced in the IDSs, which have been evaluated and compared by various classifications on the KDDCup'99 dataset. Based on the results of simulations, the statistical normalization model is introduced as the best choice for a large dataset.

Therefore, we also used statistical normalization on the dataset. The goal of statistical normalization is to convert derived data from any normal distribution to a standard normal distribution with mean zero and unit variance. Statistical normalization is defined as (1):

$$x_i = \frac{v_i - \mu}{\sigma} \qquad (1)$$

where $\mu$ is mean and $\sigma$ is its stand deviation of n values for a given feature:

$$\mu = \frac{1}{n} \sum_{i=1}^{n} v_i \qquad (2)$$

$$\sigma = \sqrt{\frac{1}{n} \sum_{i=1}^{n} (v_i - \mu)^2} \qquad (3)$$

**6. Selection of appropriate classification algorithm:** In the last step, we also evaluated the efficiency of the available classifications on the KDDCup'99 dataset in order to select the best data classification algorithm in the proposed model. The results are presented in Table VII in the next section.

## V. SIMULATION AND RESULTS

In the following, the simulation results of the proposed model on the KDDCup'99 dataset and the evaluation of different classification algorithms are presented in Table VII.

As shown in Table VII, the best detection rate and false alarms rate are respectively with 99.95% and 0.24% for the PART classification algorithm, however, the training time (0.76 seconds) and the test (0.025 seconds) is very low, which makes it perfect for use in WSNs. Therefore, we use the PART classification algorithm for the final training and test of the proposed IDS.

PART is an algorithm for inferring rules by repeatedly generating partial decision trees, thus combining the two major paradigms for rule generation: creating rules from decision trees and the separate_and_conquer rule learning technique [31].

In order to evaluate the performances of the proposed IDS, and comparing with existing works, the following criteria are considered:

- *Detection Rate:* The detection rate or the accuracy of detecting is the percentage of detected attacks relative to the total attacks.

$$Detection\ Rate = \frac{No.of\ Detected\ Attacks}{No.of\ Attacks} * 100\% \qquad (4)$$

- *False Alarm Rate:* This criterion shows an incorrect alarm rate in detecting attacks. In other words, it determines how much of the detected attacks was not attack, and the IDS mistakenly detected them.

15TABLE VII. Evaluation of different classification algorithms on the proposed model.

| | Classifiers Algorithms | TP Rate | FP Rate | Precision | F-Measure | ROC Area | Kappa Statistic | Training Time | Testing Time |
|---|---|---|---|---|---|---|---|---|---|
| **Bayes** | A1DE | 99.49 | 0.35 | 99.48 | 99.48 | 99.99 | 99.1 | 0.14 | 0.054 |
| | Bayes Net | 97.97 | 1.7 | 98.08 | 98.01 | 99.88 | 96.44 | 0.18 | 0.054 |
| | Naïve Bayes | 89.57 | 7.51 | 89.66 | 88.94 | 96.04 | 81.52 | 0.04 | 0.111 |
| | HMM | 55.43 | 55.43 | 30.72 | 39.53 | 50.00 | 0.00 | 0.12 | 0.032 |
| **Function** | MLP | 96.81 | 2.37 | 96.78 | 96.77 | 98.76 | 0.9441 | 1048.2 | 0.542 |
| | SVM | 98.11 | 1.46 | 97.98 | 98 | 98.32 | 0.9667 | 21.25 | 8.764 |
| | SMO | 93.87 | 4.8 | 93.48 | 93.54 | 95.71 | 0.8913 | 45.97 | 0.098 |
| | Logistic | 94.96 | 3.26 | 94.94 | 94.90 | 97.74 | 0.9118 | 53.55 | 0.089 |
| **Rules** | Decision Table | 98.34 | 1.39 | 98.17 | 98.25 | 99.75 | 0.9708 | 0.94 | 0.035 |
| | FURIA | 99.51 | 0.36 | 99.5 | 99.5 | 99.73 | 0.9915 | 70.19 | 0.086 |
| | JRip | 99.45 | 0.42 | 99.44 | 99.44 | 99.63 | 0.9904 | 7 | 0.031 |
| | **PART** | **99.59** | **0.24** | **99.59** | **99.58** | 99.89 | **0.9923** | 0.76 | **0.025** |
| **Trees** | J48 | 99.32 | 0.42 | 99.31 | 99.31 | 99.63 | 0.9881 | 0.55 | 0.021 |
| | Random Forest | 99.44 | 0.33 | 99.43 | 99.43 | 99.97 | 0.9901 | 6.88 | 0.433 |
| | Random Tree | 99.4 | 0.32 | 99.39 | 99.39 | 99.57 | 0.9894 | 0.12 | 0.016 |
| | REP Tree | 99.09 | 0.56 | 99.07 | 99.06 | 99.7 | 0.984 | 0.22 | 0.017 |

$$False\ positive\ Rate = \frac{No.of\ misdetected\ Attacks}{No.of\ Normal\ connections} * 100\% \tag{5}$$

- *Training time:* The time when the model is created based on training data samples.
- *Testing Time:* The time when the created model in the training step is evaluated based on the testing data samples.
- *Computational complexity:* This criterion determines the computations volume of the proposed method, which depends on the number of selected features, the number of training data samples, and the applied classification algorithm. This criterion has a direct relationship with the testing time.

TABLE VIII. The results of the comparison between the proposed IDS and the existing systems.

| # | IDS Method | Feature selected | Detection Rate | False Alarm Rate | Computational | Training Time | Testing Time |
|---|---|---|---|---|---|---|---|
| 1 | IIDS [19] | 24 | 90.96 | 2.06 | low | 135.37 | 0.29 |
| 2 | GHIDS [20] | 41 | 97.65 | 3.85 | Very high | 1229 | 73.45 |
| 3 | NHIDS [21] | 4 | 95.37 | 2.24 | low | 0.09 | 0.01 |
| 4 | ACO-SVM [22] | 25 | 98.38 | 0.004 | medium | 28.01 | 1.44 |
| 5 | GA-SVM [23] | 10 | 97.3 | 0.02 | high | 68.84 | 11.69 |
| 6 | IWD-IDS [24] | 9 | 99.41 | 1.41 | medium | 69.21 | 2.76 |
| 7 | MCFA [25] | 19 | 94.74 | 2.52 | medium | 0.84 | 1.74 |
| 8 | FCL-IDS [26] | 25 | 99.16 | 0.74 | low | 58.55 | 0.08 |
| 9 | I-NSGA-III [27] | 20 | 99.37 | 0.06 | medium | 30.2 | 1.06 |
| 10 | KBIDS [28] | 13 | 97.85 | 1.87 | medium | 84.3 | 3.83 |
| **11** | **Proposed IDS** | **4** | **99.59** | **0.24** | **low** | **0.76** | **0.025** |

In Table VIII, the results of the comparison between the proposed IDS and the existing systems are presented in terms of the criteria described above. All the results presented in the following are the average of 10 performed simulation operations. Also, for the proper comparison of the proposed method with existing works, the same dataset (KDDcup'99) whose details are presented in Table III, used to simulate all the methods.

According to the results presented in Figures 4 through 6, the proposed system with a high detection rate of 99.59% and a low false alarm rate of 0. 24%, as well as a low testing time of 0.025sec (that indicates low computational complexity), is considered as an effective and lightweight method.

As shown in Fig. 4, the detection rate of the proposed IDS is 99.59%, which has the highest rate among existing IDSs. Also, with a very low false alarm rate of 0.24%, that presented in Fig. 5, with a slight difference, is after [22], [23] and [27], but because of the high computational complexity of them, the proposed algorithm has a better condition. In addition to, the proposed IDS with a very low testing time of 0.025sec, that presented in Fig. 6, with a slight difference, is after [21], but because of the low detection rate and high false alarm rate of [21], the proposed algorithm has a better condition.

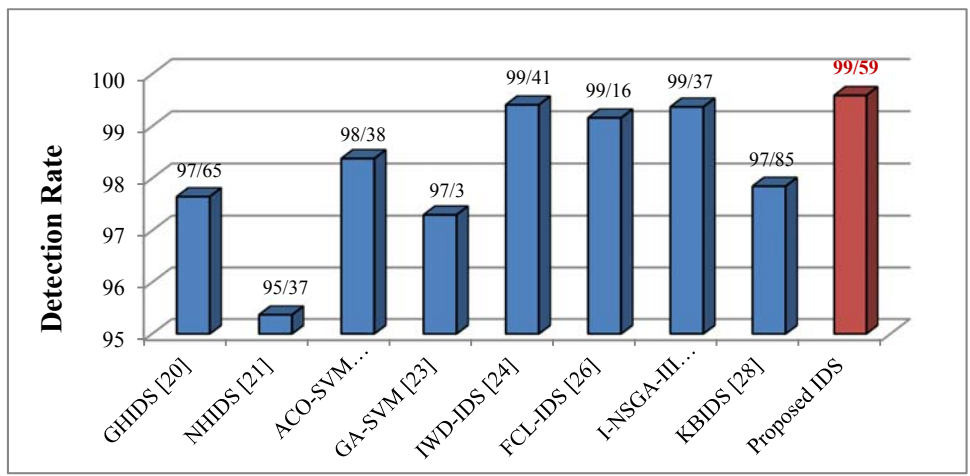

Fig. 4. Detection rate of proposed IDS and other IDSs

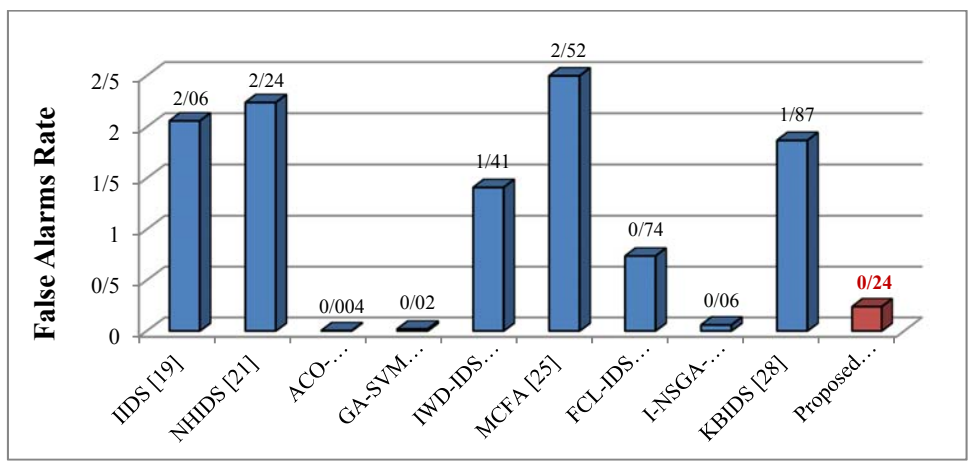

Fig. 5. False alarm rate of proposed IDS and other IDSs

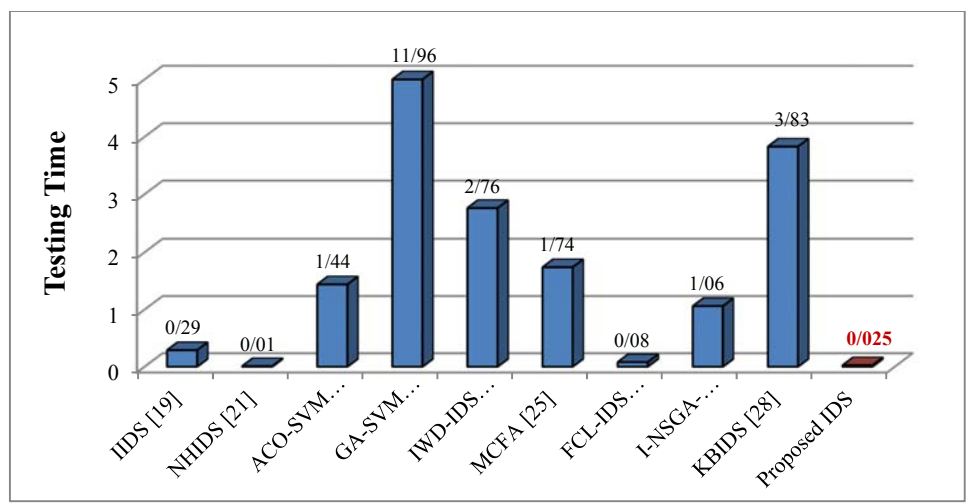

Fig. 6. Testing time of proposed IDS and other IDSs

VI. CONCLUSION

In this paper, we first introduced intrusion detection systems and then investigated various types of existing IDSs to securing CHs in WSNs. Then, considering the critical operation of the CHs, we proposed a hybrid IDS based on data mining algorithms for their security, which, using a data preprocessing model. It dramatically reduces computational complexity, and memory usage in the IDS, and provides the possibility of using classification algorithms to intrusion detection and securing CHs in WSNs. The results of the simulations show that the proposed system, in comparison with the existing ones, in addition to low computational complexity, with a high detection rate, a low false alarms rate and also a low testing time, is considered as an effective and lightweight IDS for securing CHs in WSNs.